\documentclass{elsart}
\usepackage{graphicx}

\begin{document}

\begin{frontmatter}

\title{Regular order reductions of ordinary and 
delay-differential equations}

\author[Bilbao]{J.M.\ Aguirregabiria\thanksref{JMA}},
\author[Paris]{Ll.\ Bel},
\author[Bilbao]{A.\ Hern\'andez} \and 
\author[Bilbao]{M.\ Rivas}
\address[Bilbao]{F\'{\i}sica Te\'orica, Facultad de Ciencias,
Universidad del Pa\'{\i}s Vasco, Apdo. 644, 48080~Bilbao (Spain)}
\address[Paris]{Laboratoire de Gravitation et Cosmologie Relativistes,
CNRS/URA 769, Universit\'e Pierre et Marie Curie, 4, place Jussieu.
Tour~22-12, Bo\^{\i}te~courrier~142, 75252 Paris Cedex~05 (France)}
\thanks[JMA]{Corresponding author.\\
e-mail: \texttt{wtpagagj@lg.ehu.es}\\
Tel.: +34 944647700 (ext.\ 2585)\\
FAX: +34 944648500}
 
\begin{abstract}
We present a C program to compute by successive approximations
the regular order reduction of a large class of ordinary differential
equations, which includes evolution equations in electrodynamics
and gravitation. The code may also find the regular order
reduction of delay-differential equations.
\end{abstract}

\begin{keyword}
Ordinary differential equations; Delay-differential equations;
Order reduction; Lorentz-Dirac equation; Abraham-Lorentz equation;
Chaotic scattering\\
\emph{PACS}: 02.30.Hq, 02.30.Hq, 04.50.+h, 04.90.+e, 41.60.-m, 47.52.+j\\
\emph{Program Library Index section}: 4.3 Differential Equations
\end{keyword}

\end{frontmatter}

\newpage
\textbf{PROGRAM SUMMARY}

\begin{small}

\emph{Title of program:} ODEred

\emph{Catalog identifier:}

\emph{Program obtainable from:} CPC Program Library, 
Queen's University of Belfast, N.\ Ireland

\emph{Licensing provisions:} none

\emph{Computers:} The code should work on any computer
with an ANSI C compiler. It has been tested on a PC,
a DEC AlphaServer 1000 and a DEC Alpha AXP 3000-800S.

\emph{Operating system:} The code has been tested under 
Windows 95, Digital Unix v.4.08 (Rev. 564) and Open VMS V6.1.

\emph{Programming language used:} ANSI C

\emph{Memory required to execute with typical data:} It depends
on the number of equations and retarded points stored
in memory. Many intersting problems can be solved in 1 Mbyte.

\emph{No. of bytes in distributed program, including test data, etc.:}
84.695

\emph{Distribution format:} ASCII

\emph{Keywords:}
Ordinary differential equations; Delay-differential equations;
Order reduction; Lorentz-Dirac equation; Abraham-Lorentz equation;
Chaotic scattering

\emph{Nature of the physical problem}\\
In different physical problems, including electrodynamics
and theories of gravitation, there appear singular differential
equations whose order decreases when a physical parameter takes
a particular but very important value. Typically most
solutions of these equations are unphysical. 
The regular order reduction is an equation of lower order which contains
precisely the physical solutions, which are those regular in
that parameter. The program computes the solution of the regular order 
reduction for
a large set of ordinary and delay-differential equations.

\emph{Method of solution}\\
The basic integration routine is based on the continuous 
Prince-Dormand method of eighth order.
At each integration step, successive approximations are performed by
using the polynomial interpolating the solution that has been computed
in the previous approximation.

\emph{Typical running time}\\
It depends heavily on the number and complexity of the equations and
on the desired solution range. It was at most
a couple of seconds in the test problems.
%

\end{small}

\newpage
\textbf{LONG WRITE-UP}

\section{Introduction}

In different physical theories there appear singular evolution equations
that share some common properties: most of their solutions are
unphysical because their order is higher than expected except for a
particular, but important, value of a parameter for which the order
reduces to what one would expect on physical grounds. For instance, the
Lorentz-Dirac equation~\cite{Rohrlich} that describes the motion of a
charged particle with radiation reaction is of third order, so that
initial position and velocity would not be enough to determine its
evolution and most of its mathematical solutions are `runaway', i.e. the
acceleration increases without limit. Nevertheless, in the limit when
the charge goes to zero the Lorentz-Dirac equation becomes a
second-order equation while the non physical solutions diverge for that
value of the parameter.

Equations of this kind show strong numerical instabilities, which
prevent from integrating them forward in time: even if the physical
initial conditions are chosen, the integration error introduces an
initially small contribution from the non physical solutions which then
blow out~\cite{Huschilt,Baylis}. The standard recipe to avoid this
problem is to integrate  backwards~\cite{Huschilt,Baylis}, but this is
impossible in many cases because the final state from which to integrate
is unknown~\cite{JMA}.

A natural approach to this kind of problems is provided by the concept 
of ``regular order reduction'', which is an evolution equation with the 
right order that contains precisely the physical solutions and behaves
smoothly for the particular value of the parameter for which the order
of the singular equation decreases~\cite{Bhabha}. In the context of the
Lorentz-Dirac equation this concept was discussed by
Kerner~\cite{Kerner} and Sanz~\cite{Sanz}. Order reductions have been
also used to replace the delay-differential equations that appear in the
electrodynamics~\cite{Bel2,Bel3} and in non-linear optics~\cite{Bel4},
as well as to analyse fourth-order equations that appear in theories of
gravitation with a quadratic Lagrangian~\cite{Bel1} and in the study of
quantum corrections to Einstein equations~\cite{Parker}. 

Except in rather trivial cases, the order reduction cannot be computed 
exactly and some approximation scheme is necessary. One may use
a power expansion~\cite{Sanz}, but the explicit expressions become quickly
too complex. Several years ago, one of us (Ll.B.) wrote a routine
to find the regular order reduction of delay-differential equations. 
To compute numerically the order reduction of singular ordinary
differential equations we have proposed and analysed in some cases a 
method of successive approximations~\cite{JMA,sanedrin}. 

The goal of this paper is twofold: we want to make widely available the
code for the latter method, which we developed first for a very 
particular system~\cite{ODE}, and to discuss its applicability in a case
that was not analysed previously: that of delay-differential equations.

\section{The general problem}

Let us assume that an evolution equation
can be written (after inserting additional variables and trivial 
equations to make it a first-order system) in the form of $n$ 
differential equations that in vectorial notation appear as
\begin{equation}
\dot\mathbf{x} = 
\mathbf{f}\left(t,\mathbf{x},\dot\mathbf{x},\ddot\mathbf{x},\ldots\right),
\label{eq:system}
\end{equation}
and that we suspect on physical grounds that the successive
approximations
\begin{eqnarray}
\dot\mathbf{x} &=& 
\mathbf{f}_0\left(t,\mathbf{x}\right),\label{eq:apprxo0}\\
&\vdots&\nonumber\\
\dot\mathbf{x} &=& 
\mathbf{f}_{n+1}\left[t,\mathbf{x},
\dot\mathbf{g}_n\left(t;t_0,\mathbf{x}_0\right),
\ddot\mathbf{g}_n\left(t;t_0,\mathbf{x}_0\right),\ldots\right],
\qquad(n=0,1,\ldots),
\label{eq:apprxon}
\end{eqnarray}
converge in some domain when one uses an appropriate starting 
point~(\ref{eq:apprxo0}) that is suggested by the physical problem.
In many cases, but not necessarily, all the $\mathbf{f}_n$ but the first one
will have the same functional structure.
The key point is that at each stage a first-order system has to be
solved because higher derivatives appearing in~(\ref{eq:system})
are missing in~(\ref{eq:apprxo0}) and have been replaced 
in~(\ref{eq:apprxon}) by the (numerical) solution
$\mathbf{g}_{n}\left(t;t_0,\mathbf{x}_0\right)$ 
computed in
the previous approximation:
\begin{eqnarray}
\dot\mathbf{g}_0 &=& 
\mathbf{f}_0\left(t,\mathbf{g}_0\right),\qquad\qquad\qquad\qquad\ \ %
\mathbf{g}_0\left(t_0;t_0,\mathbf{x}_0\right)=\mathbf{x}_0,
\label{eq:sol0}\\
\dot\mathbf{g}_n &=&
\mathbf{f}_n\left(t,\mathbf{g}_n,\dot\mathbf{g}_{n-1},
\ddot\mathbf{g}_{n-1},\ldots\right),\qquad
\mathbf{g}_n\left(t_0;t_0,\mathbf{x}_0\right)=\mathbf{x}_0.
\label{eq:soln}
\end{eqnarray}

For simplicity we have assumed above that the differential equations
are ordinary, i.e. that $\mathbf{x}$ and its derivatives
were evaluated at a single time $t$. But nothing prevents us for
considering more general cases in which some values appearing
on the right hand side are evaluated at one or several earlier times. 
In such
a case we have a delay-differential system and
its phase-space is infinite-dimensional because 
the initial condition is not a single $n$-dimensional point 
$\mathbf{x_0}$ but a full vector function $\mathbf{h}(t)$ 
for some appropriate interval $t_{-1}\le t\le t_0$~\cite{Bellman}.
We will consider an example of this type in section~\ref{sec:dde}.

Except in some special cases~\cite{JMA,sanedrin}, nothing is known about
the convergence of this scheme of succesive approximations and we are
not claiming that the method will converge for all (or at least a large
class of) mathematical problems of this kind, but that we think that it
will work for many interesting physical problems. (We refer
to~\cite{Bel3,Bel4,JMA,sanedrin} for the physical motivations  of our
approach.) In fact one of the main reasons to write the code we will
describe in the  next section was that it can be used to explore
numerically in which cases, and for which domains, the convergence is
likely to happen. One should also keep in mind that the numerical
results of the method of successive approximations can be checked
\emph{a posteriori} by solving the singular ordinary differential
equation backwards from the final state computed by means of the code
we describe in the next section.

\section{The code}

Our code attempts to construct automatically a good approximation
to the result obtained in the limit, i.e. to the solution of the
regular order reduction, by constructing the successive 
approximations~(\ref{eq:soln}) at each integration step. 

Since the solution $\mathbf{g}_n$ appearing on the right hand side 
of~(\ref{eq:apprxon}) must be evaluated, along with its derivatives,
for different values of $t$ in the interval corresponding to
the current integration step, we need not only the values of
$\mathbf{g}_n$ on the initial and final points of the interval
but a method to compute it along the whole interval. Although to
test our ideas we developed first a method based on the
classical fourth-order Runge-Kutta algorithm with Hermite
interpolation, we present here only
the final code based on the eighth-order Prince-Dormand method 
implemented by Hairer, Norsett and Wanner~\cite{Hairer}, as well
as an alternative fifth-order code based on the
Dormand-Prince method~\cite{Hairer}, which can be used for
testing purposes.

Both codes are embedded Runge-Kutta methods with automatic step-size
control that provide not only the values of the solution at discrete
points but also a polynomial which interpolates the solution between
each pair of points, allowing us to compute the desired values and
derivatives. Of course, one cannot expect a good accuracy if derivatives
of high order are necessary, but we have found in practice that, even in
difficult chaotic cases, the seventh-order polynomial of the
eighth-order method is good enough in many interesting cases where only
first and second derivatives are needed.  The second method provides a
fourth-order interpolating polynomial which can be used if only low
accuracy results are required. The availability of the interpolating
polynomial is also the key feature needed to deal with
delay-differential equations.

Our codes are loosely based on the routines \texttt{DOP853} and
\texttt{RETARD} by Hairer, Norsett and Wanner~\cite{Hairer}. They
consist of four public functions that are declared in \texttt{ODEred.h},
which you have to include in your code to take advantage of ANSI
declarations. You have to compile and link with your code the file
containing the implementation of these functions: \texttt{ODEred.c} in
the case of the eighth-order method and \texttt{ODEred5.c} for the
fifth-order code.

\subsection{Public function \texttt{InitODEred}}

Before using the main integration routine (or when one of the three
parameters below changes) the following call must be made to initialise
some global values used to store the solution even after the integration
routine exits:

\texttt{r = InitODEred(issyst,dimen,points);}

\subsubsection{Input parameters}
\begin{description}
\item[\texttt{issyst}] An integer that should be always 
equal to \texttt{TRUE} (nonzero), except when the system of $n$ first-order
equations is equivalent to a single equation of order $n$ because
the first $n-1$ equations are just in the form $\dot x_i=x_{i+1}$.
In the latter case, \texttt{issyst} may be set to \texttt{FALSE} (0). 
Although an equation must be written with this package in the form 
of an equivalent system, this value controls the meaning 
of the first parameter in \texttt{InterpolateODEred} 
(see section~\ref{sec:interpol}), which may be used in a 
more natural way with this option. 
\item[\texttt{dimen}] A positive integer with the number $n$ of 
first-order equations. This is the dimension of the regular order
reduction.
\item[\texttt{points}] A positive integer with the number of solution points 
to be stored in memory. In the case of ordinary differential equations 
1 will be enough because the code only needs the last point to compute
interpolated values and derivatives. If the equations contain delay, 
it is necessary to compute values for retarded values of the independent
variable, so the corresponding polynomial coefficients must be still in 
memory. Since the code chooses automatically the step size, it is not easy
to know beforehand how many points are necessary for a given accuracy,
especially if the delay is not constant. In practice one can start
with a value of 100 and make it higher if the \texttt{InterpolateODEred}
described below returns the error code -1.
\end{description}

\subsubsection{Return value}
An integer \texttt{r} equal to -1 if the initialization failed by 
lack of memory, and to \texttt{points} if it was successful.

This routine and \texttt{EndODEred} could have been included inside
\texttt{SolveODEred}, but we prefer this approach, because
in some cases it is convenient to make more than one call to 
\texttt{SolveODEred} with the same stored solution, which moreover 
often has to be used even after \texttt{SolveODEred} exits.

\subsection{Public function \texttt{SolveODEred}}

The actual numerical integration of the system is performed by using:

\begin{verbatim}
r = SolveODEred(&t,x,t0,tend,equation,output,
                eps,hmax,h,accuracy,maxiter,forget);
\end{verbatim}
\subsubsection{Input parameters}
\begin{description}
\item[\texttt{t}] Initial value for $t$. Passed by reference. 
\item[\texttt{x[i]}] Initial values for $\mathbf{x}$. Passed by reference. 
\item[\texttt{t0}] Final $t$-value for initial conditions in retarded systems:
the initial function \texttt{initial} (see section~\ref{sec:init}) 
will be used 
for $t < \mathtt{t0}$, if it is different from \texttt{NULL}. 
It is not automatically made equal to the
initial value of $t$ because sometimes \texttt{SolveODEred} is called
more than just once with the same \texttt{t0} but different $t$s (for
instance, when the integration is paused and then continued). 
\item[\texttt{tend}] Final $t$-value. \texttt{tend}-\texttt{t} may be 
negative (except for delay-differential systems) to integrate backwards.
\item[\texttt{equation}] A user-provided function returning the 
derivatives on the right hand side of the equations in the system
(see section~\ref{sec:equa}).
\item[\texttt{output}] The output function to be called at each solution point
(see section~\ref{sec:outp}).
\item[\texttt{eps}] Local tolerance for step-size control.
\item[\texttt{hmax}] Maximum step-size.
\item[\texttt{h}] Initial step-size guess.
\item[\texttt{accuracy}] Maximum relative error among successive approximations.
\item[\texttt{maxiter}] Maximum number of iterations.
\item[\texttt{forget}] If this integer is \texttt{TRUE} previously computed
solution points will be removed from memory.
\end{description}

\subsubsection{Output parameters}
\begin{description}
\item[\texttt{t}] Last value for $t$. Passed by reference.
\item[\texttt{x[i]}] Final values for $\mathbf{x}$. Passed by reference.
\end{description}

\subsubsection{Return value}
An integer equal to 0 on success, to -1 if while trying to keep the 
truncation error under \texttt{eps} the step size became too small,
-10 if the maximum number of iterations \texttt{maxiter}
was performed without attaining the desired \texttt{accuracy}.
Positive values are error codes returned by \texttt{output}.

\subsection{Public function \texttt{InterpolateODEred}}
\label{sec:interpol}
The user-provided functions \texttt{equation} and
\texttt{output} may compute the values of the solution
for any value of $t$ by using

\texttt{r = InterpolateODEred(n,t,x,initial);}

As an added feature, it is also able to compute derivatives of the
solution.

\subsubsection{Input parameters}
\begin{description}
\item[\texttt{n}] Integer indicating the ``component'' to be computed. 
If $\mathtt{n}\ge0$, the \texttt{n}-th component 
will be returned. If $\mathtt{n} < 0$, the components 
$0,\ldots,|\mathtt{n}|-1$ will be computed.\\
For $j = 0,\ldots,\mathtt{dimen}-1$,
the $j$-th component is always the $j$-th dependent variable $x_j$ 
(i.e. the $j$-th component of $\mathbf{x}$).\\ 
If \texttt{issyst} is \texttt{TRUE}, the index 
$\mathtt{n} = \mathtt{dimen}\times k+j$ refer to the $k$-th 
derivative of component $x_j$, with $j = 0,\ldots,\mathtt{dimen}-1$
and $k=1,2,\ldots$\\
If \texttt{issyst} is \texttt{FALSE}, the system must be
equivalent to a single equation of orden $\mathtt{dimen}$, in such
a way that the dependent variable $x_j$ is the derivative of component
$x_{j-1}$ for $j = 1,\ldots,\mathtt{dimen}-1$.
In this case, the index 
$\mathtt{n} = \mathtt{dimen}-1+k$ corresponds to the $k$-th 
derivative of component $\mathtt{dimen}-1$ and is, thus, the
$\mathtt{n}$-th derivative of the first dependent variable $x_0$. 
\item\texttt{t} Independent variable for which the value(s) must be computed.
\item\texttt{initial} This parameter will be \texttt{NULL} in general,
but can be a function defined to return from the call \texttt{initial(n,t)}
with the \texttt{n}-th component of the initial function in retarded systems
for values $\mathtt{t} < \mathtt{t0}$. See section~\ref{sec:outp}.
\end{description}

\subsubsection{Output parameter}
\begin{description}
\item[\texttt{x}] Pointer to a double, or, 
if $\mathtt{n} < 0$, to an array with at least $|\mathtt{n}|$ doubles, where
the interpolated value(s) will be stored. 
\end{description}

\subsubsection{Return value}
An integer equal to 0 on success, to -1 if \texttt{t} is out of the range stored in
memory (or available through \texttt{initial}), and to
-2 if \texttt{n} (or $|\mathtt{n}|-1$, if $\mathtt{n} < 0$) is greater than the
maximum value.

\subsection{Public function \texttt{EndODEred}}
When the ODEred package is no longer necessary, the program must use

\texttt{EndODEred();}

to release the allocated memory.

On the other hand, the user must provide the following two functions
when \texttt{SolveODEred} is called. 

\subsection{User-provided function \texttt{equation}}
\label{sec:equa}
When calling \texttt{SolveODEred} one has to provide a pointer
to a function that, whatever its name is, has the following structure:
\begin{verbatim}
double equation(int     n,     /* Iteration number     */
                double  t,     /* Independent variable */
                double *x,     /* Dependent variables  */
                double *f)     /* Returned derivatives */
{
  ...
  f[...] = ...;
  ...
}
\end{verbatim}
It defines the differential system and returns 
$\mathbf{f}_n\left(t,\mathbf{x},\dot\mathbf{g}_{n-1},
\ddot\mathbf{g}_{n-1},\ldots\right)$. 
The function knows the $\mathbf{f}_n$ it has to compute by means of 
the parameter \texttt{n} and, if $\mathtt{n} > 0$ uses 
\texttt{InterpolateODEred}
to find the $\mathbf{g}_{n-1}$ values (and their derivatives),
 which were computed in the previous approximation. In the
example of section~\ref{sec:test1} one could use the following
function:
\begin{verbatim}
void equation(int iteration,double t,double *x,double *f)
{
  if (iteration == 0) {       
    *f = -A0*(*x);
  }
  else {
    double x2;
    InterpolateODEred(2,t,&x2,NULL);
    *f = -A0*(*x)+EPSILON*x2;
  }
}
\end{verbatim}
The first time the routine is called at each step \texttt{iteration} is
null and a first-order equation is solved. In the remaining iterations
the interpolating polynomial is used through the function
\texttt{InterpolateODEred} to compute the value of the second
derivative appearing in the right-hand-side of the complete equation.

\subsection{User-provided function \texttt{output}}
\label{sec:outp}
When calling \texttt{SolveODEred} one has to provide a pointer
to a function with the following declaration (and any name):
\begin{verbatim}
int output(int     n,  /* Iteration number     */
           double  t,  /* Independent variable */
           double *x)  /* Dependent variables  */
\end{verbatim}
Each time an integration step has been performed
\texttt{SolveODEred} calls this function with the current values 
of the independent and dependent variables. 
This allows using the computed solution (to print it,
for instance). 
The solution and its derivatives for other values of $t$ 
may be found by means of \texttt{InterpolateODEred}. 
The parameter \texttt{n} may be used to know how many successive 
approximations were necessary to attain the desired \texttt{accuracy}. 
The function must return a status code in a nonnegative integer:
if it is different from 0, \texttt{SolveODEred} will stop immediately
and return this value as an error code. This allows exiting the integration 
when some condition is met even before \texttt{tend} is reached
(see the example in section~\ref{sec:test2}).

\subsection{User-provided function \texttt{initial}}
\label{sec:init}
When dealing with delay-differential equations one may define a pointer
to a function which returns the initial functions for $t < \mathtt{t0}$.
The function is declared as
\begin{verbatim}
double initial(int     n,  /* Component index      */
               double  t)  /* Independent variable */
\end{verbatim}
may have any name, and must return the value of the \texttt{n}-th 
component of the dependent variable $\mathbf{x}$ for
$t=\mathtt{t}$. This possibility is provided for generality, so that
the routine could be used to integrate delay-differential equations, but
notice that in this case the package should be used with 
\texttt{maxiter} = \texttt{accuracy} = 0, which 
disables the mechanism of successive approximations. 
To compute the
regular order reduction of a delay-differential equation one 
must not provide this function because the algorithm will not
converge unless it happens to be the one corresponding to the
as yet unknown reduction. Calling \texttt{SolveODEred}
with a \texttt{NULL} pointer will instruct the algorithm to use
successive approximations to find the initial functions from
the initial values $\mathbf{x}_0$.

\subsection{The algorithm}

The algorithm is a direct implementation 
of~(\ref{eq:apprxo0})--(\ref{eq:apprxon}) applied 
at each integration step, rather than to the whole integration range. 
First
the next point is computed with a Prince-Dormand step by
using $\mathtt{n} = 0$ when calling \texttt{equation}, which
effectively solves~(\ref{eq:apprxo0}).
If the integration error is estimated to be below the relative error 
in \texttt{eps}~\cite{Hairer}, the coefficients of the interpolating
polynomial are stored
and the whole step is repeated with $\mathtt{n} = 1$, and so
on until one of the following things happens:
\begin{enumerate}
\item If the relative error between the last two approximations
to the next solution point is below \texttt{accuracy}, the step
is accepted, because this is the desired approximation
to the solution of the regular order reduction in the step. This
estimation of the relative error (along with the automatic 
step-size control) allows getting the desired approximation with the
minimum computational effort. But one can also compute the
same number of approximations at every step, by disabling 
\texttt{accuracy}, as explained now.
\item If \texttt{accuracy} is 0 and \texttt{maxiter} iterations 
have been computed, the step is accepted. This is useful
to compute in all steps a fixed number of approximations to explore how the
algorithm works in different problems, or if one knows that
further approximations cannot meaningfully improve the solution because the
original equation was itself approximate~\cite{Parker}.
\item If \texttt{accuracy} is nonzero and \texttt{maxiter} 
iterations have been performed without getting a value for
the relative error between approximations below \texttt{accuracy}, 
\texttt{SolveODEred} immediately
returns with an error code equal to -10. This may happen
because \texttt{accuracy} and/or \texttt{maxiter} are too low
or because we are out of the domain in which the method
of successive approximations converges.
\end{enumerate}
In the first two cases above the coefficients in the interpolating
polynomial are stored in another location, so that the information
corresponding to the last \texttt{points} is available through
\texttt{InterpolateODEred} to integrate delay-differential equations 
and to make possible continuous output
in the \texttt{output} function, which is then invoked
to allow the calling program using the new solution point. 
If \texttt{output} does not return a positive error code and
\texttt{tend} has not been reached, one continues with the next step.

If at any point the integration error is estimated to be above the relative 
error in \texttt{eps}, all the data corresponding to the current
step is discarded and one starts again with $\mathtt{n} = 0$ and
a smaller value for the step-size computed according with
the strategy discussed in~\cite{Hairer}. 
In the same way, the step-size is automatically increased when the
estimated truncation error is too low.
If the step-size becomes
too small for the working (double) precision, \texttt{SolveODEred}
returns with an error code equal to -1.

If both \texttt{accuracy} and \texttt{maxiter} are null, the successive
approximations are disabled and \texttt{SolveODEred} is just an 
ODE integrator.

\section{Test cases}

The code has been checked in a variety of cases by using
a GUI environment designed to analyse dynamical systems~\cite{ODE}. 
Some of them have been discussed
elsewhere~\cite{JMA,sanedrin} and we present here some new
examples for which the driver programs are provided.

\subsection{The simplest example}
\label{sec:test1}

The regular order reduction of the equation
\begin{equation}
\dot x=-a_0 x+\epsilon \ddot x.
\label{eq:dotddotx}
\end{equation}
is 
\begin{equation}
\dot x=-a x, \qquad a \equiv \frac{\sqrt{1+4a_0\epsilon}-1}{2\epsilon}
\label{eq:dotx}
\end{equation}
and it can be seen by an argument similar to the one used in~\cite{JMA}
that the analytical method of succesive 
approximations~(\ref{eq:apprxo0})--(\ref{eq:apprxon}) will converge to 
it for $0\le a_0\epsilon<3/4$.
Since we know the solution, $x=x_0 \e^{-at}$, of the regular order
reduction~(\ref{eq:dotx}) for the second-order 
equation~(\ref{eq:dotddotx}), we can directly check the global
integration error in this simple case. In the provided 
\texttt{TestRed1.c} file, the problem is solved for $a_0=1$,
$\epsilon=0.1$ and $x(0)=1$ (a single initial condition!) with 
\texttt{eps} = $10^{-10}$, \texttt{hmax} = 1,
\texttt{accuracy} = $10^{-8}$ and
\texttt{maxiter}  = $100$.
To illustrate the result we collect in the TEST RUN OUTPUT section
some solution points 
---which have been interpolated with \texttt{InterpolateODEred}---
as well as the accumulated global error, 
$(x_\mathrm{numerical}-x_\mathrm{exact})/|x_\mathrm{exact}|$,
and the number of iterations the routine needed to reach the
required \texttt{accuracy}. The evolution of this error, which
essentially measures the performance of the interpolating polynomial, 
is displayed in figure~1. When using the eighth-order method the 
actual integration step is rather long, about 0.45, and each 
interpolating polynomial is used for a wide domain, which explains 
the relatively large variation of the error over a single step. 
In the same figure we see
that if one lowers the maximum step size by setting \texttt{hmax} = 0.1
the error behaves more smoothly and is about one order of magnitude lower.
For comparison, we have included the result when the fifth order
routine is used: the error is clearly higher and far more steps 
are needed, so that one obtains the same results with 
\texttt{hmax} = 0.1 and 1. We have found in practice that this
happens in very different problems: the fifth-order method can be
used only if modest accuracy is required and the eighth-order
method and the associated interpolating polynomial behave rather
well, especially if after some essays one guess the right value
for \texttt{hmax}.

\subsection{Chaotic scattering and Abraham-Lorentz equation}
\label{sec:test2}

Of course, the code would not be useful if it were able to deal
only with linear equations. To check it under far more difficult
conditions we will consider the chaotic scattering of
a charged particle with radiation reaction moving in a plane 
around four
other equal particles held at rest in the vertices $\mathbf{x}_i$ 
of a square (see figure~2). The equation of
motion is in this case the Lorentz-Dirac equation~\cite{Rohrlich}
and the problem has been analysed first in~\cite{JMA}. Here we will
consider a simplified case in which we use the non-relativistic
approximation given by the Abraham-Lorentz equation, which
is in this case:
\begin{equation}
\frac{\d^2\mathbf{x}}{\d t^2} = 
\frac{e^2}{m}\sum_{i=1}^4{\frac{\mathbf{x}-\mathbf{x}_i}{|\mathbf{x}-\mathbf{x}_i|^3}+
\tau_0\frac{\d^3\mathbf{x}}{\d t^3}},\qquad\tau_0\equiv\frac{2e^2}{3mc^3}.
\label{eq:AL}
\end{equation}
This is a singular
third-order equation, which reduces to a second-order equation
both when the particle charge $e\to0$
and when one neglects the radiation reaction ($\tau_0\to0$). Any
of these two
second-order equations can be used as starting point for the 
succesive approximations, because if one starts from the
free case, $\ddot \mathbf{x}=0$, the next approximation
is just the second-order equation obtained by taking $\tau_0=0$
in~(\ref{eq:AL}).
For plane motion, one has to consider four first-order equations
for the derivatives of the position and velocity 
components. 
In figure~2 we see the first approximations
one gets with \texttt{TestRed2.c} by setting $\mathtt{accuracy} = 0$ and 
$\mathtt{maxiter} = 0,1,2,3,4$, as well as the approximation
to the regular reduction obtained with $\mathtt{accuracy} = 10^{-10}$,
which is labelled as AL and needs between 4 and 19 iterations depending on the charge
position and its distance to the fixed charges.

In this case we do not know the exact solution, but we can check the 
quality of the solution computed by our code by integrating backwards
from the end point towards the starting one by means of any ordinary
differential equation solver, because the unphysical modes are
exponentially damped out when integrating backwards~\cite{Huschilt}.
Integrating backwards with \texttt{ODEred} ---which can be used as a
good solver for ordinary and delay differential equations--- is around
ten times faster than integrating forward in this problem, but notice
that trying to guess the final state from which to integrate backwards
to the desired initial conditions would be impossible in a capture
process in which nothing is known a priori about the final state, and
very costly in the example we are analysing due to the sensitive
dependence on initial conditions associated to chaos and the fact that
the final velocity may be very different (around 25\% lower in the
trajectory displayed in figure~2) from the initial one due to the
radiation. 
In \texttt{TestRed2.c} to  perform this backward integration one simply
applies the same  \texttt{SolveODEred} (with \texttt{maxiter} =
\texttt{accuracy} = 0 to disable the successive approximations) to the
six first-order equations necessary to write the third-order system in a
plane. The initial conditions now include the derivative of the
acceleration, which is computed by \texttt{InterpolateODEred} from the
numerical solution constructed to approximate the regular order
reduction in the forward integration. The result of this backward
integration is also displayed in figure~2,  where it is
indistinguishable from the forward integration AL in which the code 
constructed the order reduction. The detailed error,
$|x_\mathrm{forward}-x_\mathrm{backward}|/(|x_\mathrm{forward}|+|x_\mathrm{backward}|)$,
is displayed in figure~3.

\subsection{A delay-differential equation}
\label{sec:dde}

Our code is also able to deal
with delay-differential equations in which some arguments appear 
evaluated at retarded values of the independent variable. To check
the algorithm in this context we will consider the following 
linear but illustrative example:
\begin{equation}
\dot x(t) = -a x(t-r),\qquad\mathrm{with\ }a,r >0,
\label{eq:delay}
\end{equation}
where, in fact, the only dimensionless parameter is $ar$. 
(More interesting but
far more complex retarded equations appear in  non-linear
optics~\cite{Bel4} and electrodynamics~\cite{Bel2,Bel3}.)
This is
an infinite-dimensional system because the initial
condition is a full function defined in $\left[t_0-r,t_0\right]$
but it is singular and becomes a first-order
ordinary equation both when $a\to0$ and $r\to0$. We are interested in the
corresponding regular order reduction, which happens to be
\begin{equation}
\dot x(t) = \frac{W(-ar)}{r}x(t),
\label{eq:nodelay}
\end{equation}
where $W$ is the principal branch of the Lambert W function, which is 
defined implicitly in $W \e^W=z$ as a 
generalization of the logarithm~\cite{lambert}. 
Furthermore, by using the fact that 
$z^{z^{z^{^{.^{.^.}}}}} = -W(-\ln z)/\ln z$~\cite{lambert}, 
it can be seen that the 
succesive 
approximations~(\ref{eq:apprxo0})--(\ref{eq:apprxon}) will converge 
to~(\ref{eq:nodelay}) for $0\le ar<\e^{-1}$. One can check this
by using \texttt{TestRed3.c}, 
as shown in the TEST RUN OUTPUT and in figure~4, for 
$a=1$, $r=0.3$, $x(0)=1$.
We want to stress that since at each approximation an ordinary differential
function has to be solved, no \texttt{initial} function has to be
provided but only the initial value.

\ack

The work of J.M.A., A.H. and M.R. has been partially supported by The University of 
the Basque Country under contract UPV/EHU 172.310-EB036/95.
J.M.A. and M.R have also been supported by DGICYT project PB96-0250.

\newpage
\section*{TEST RUN OUTPUT}

\subsection*{Output from \texttt{TestRed1.c}}

\begin{verbatim}
t = 1 x = 0.400084  Error = -9.65801e-07 Iterations = 15
t = 2 x = 0.160067  Error = -2.36572e-06 Iterations = 15
t = 3 x = 0.0640403 Error = -4.06969e-06 Iterations = 15
t = 4 x = 0.0256215 Error = -6.01646e-06 Iterations = 15
t = 5 x = 0.0102507 Error = -8.41758e-06 Iterations = 14
\end{verbatim}

\subsection*{Output from \texttt{TestRed3.c}}

\begin{verbatim}
t = 0.1      x = 0.849485    Error = 7.31759e-05 Iterations = 46 
t = 0.397345 x = 0.523003    Error = 4.58862e-05 Iterations = 48 
t = 0.704736 x = 0.316763    Error = 4.47788e-05 Iterations = 48 
t = 1.02387  x = 0.188211    Error = 4.35142e-05 Iterations = 48 
t = 1.35798  x = 0.109132    Error = 4.19486e-05 Iterations = 48 
t = 1.7106   x = 0.0613964   Error = 4.00809e-05 Iterations = 48
t = 2.08563  x = 0.0333008   Error = 3.79189e-05 Iterations = 48
t = 2.48743  x = 0.0172902   Error = 3.54728e-05 Iterations = 48
t = 2.92103  x = 0.00852357  Error = 3.27512e-05 Iterations = 48 
t = 3.39243  x = 0.00395061  Error = 2.97599e-05 Iterations = 48 
t = 3.90905  x = 0.00170085  Error = 2.65055e-05 Iterations = 48 
t = 4.48039  x = 0.000669746 Error = 2.30013e-05 Iterations = 48
t = 5        x = 0.000286944 Error = 2.63023e-05 Iterations = 48
\end{verbatim}

\newpage
\section*{FIGURE CAPTIONS}

Figure 1. Evolution of the accumulated error when computing
by means \texttt{TestRed1.c}
the regular order reduction~(\ref{eq:dotx}) of~(\ref{eq:dotddotx}).

Figure 2. Solutions of~(\ref{eq:AL}) corresponding to 
$\mathtt{maxiter} = 0,1,2,3,4$ with $\mathtt{accuracy} = 0$. 
The solution of
the regular order reduction of~(\ref{eq:AL}) and the
one obtained integrating it backwards are indistinguishable
and appear labelled as AL.

Figure 3. Evolution of the relative distance between the solutions
obtained integrating forward and backward system~(\ref{eq:AL})
by means of \texttt{TestRed2.c}.

Figure 4. Evolution of the accumulated error when computing
by means \texttt{TestRed3.c}
the regular order reduction~(\ref{eq:nodelay}) of~(\ref{eq:delay}).

\newpage
\vspace*{2cm}
\begin{center}
\includegraphics[width=\textwidth]{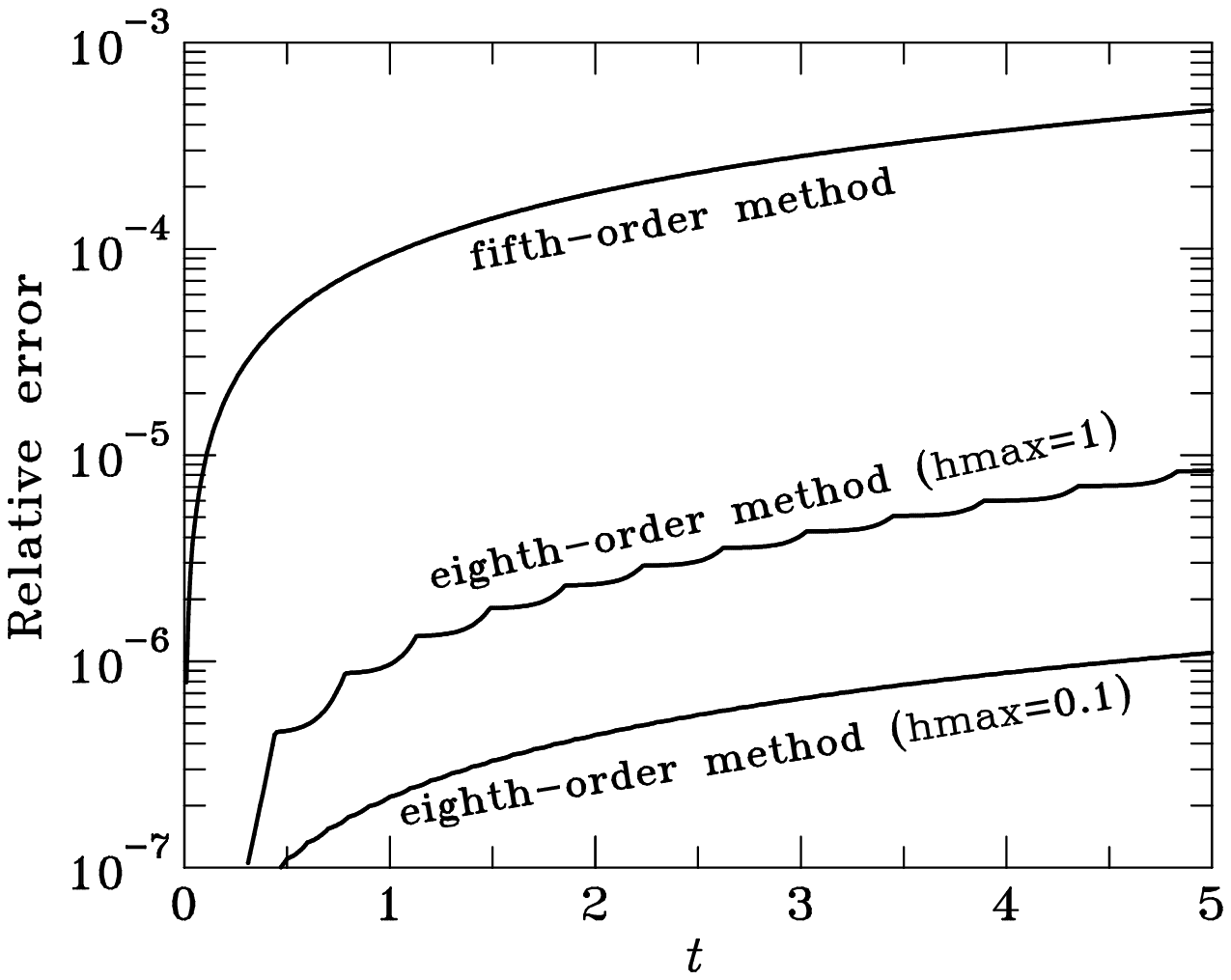}
\end{center}

\vspace{2cm}
Figure 1. Evolution of the accumulated error when computing
by means \texttt{TestRed1.c}
the regular order reduction~(\ref{eq:dotx}) of~(\ref{eq:dotddotx}).

\newpage
\vspace*{2cm}
\begin{center}
\includegraphics[width=\textwidth]{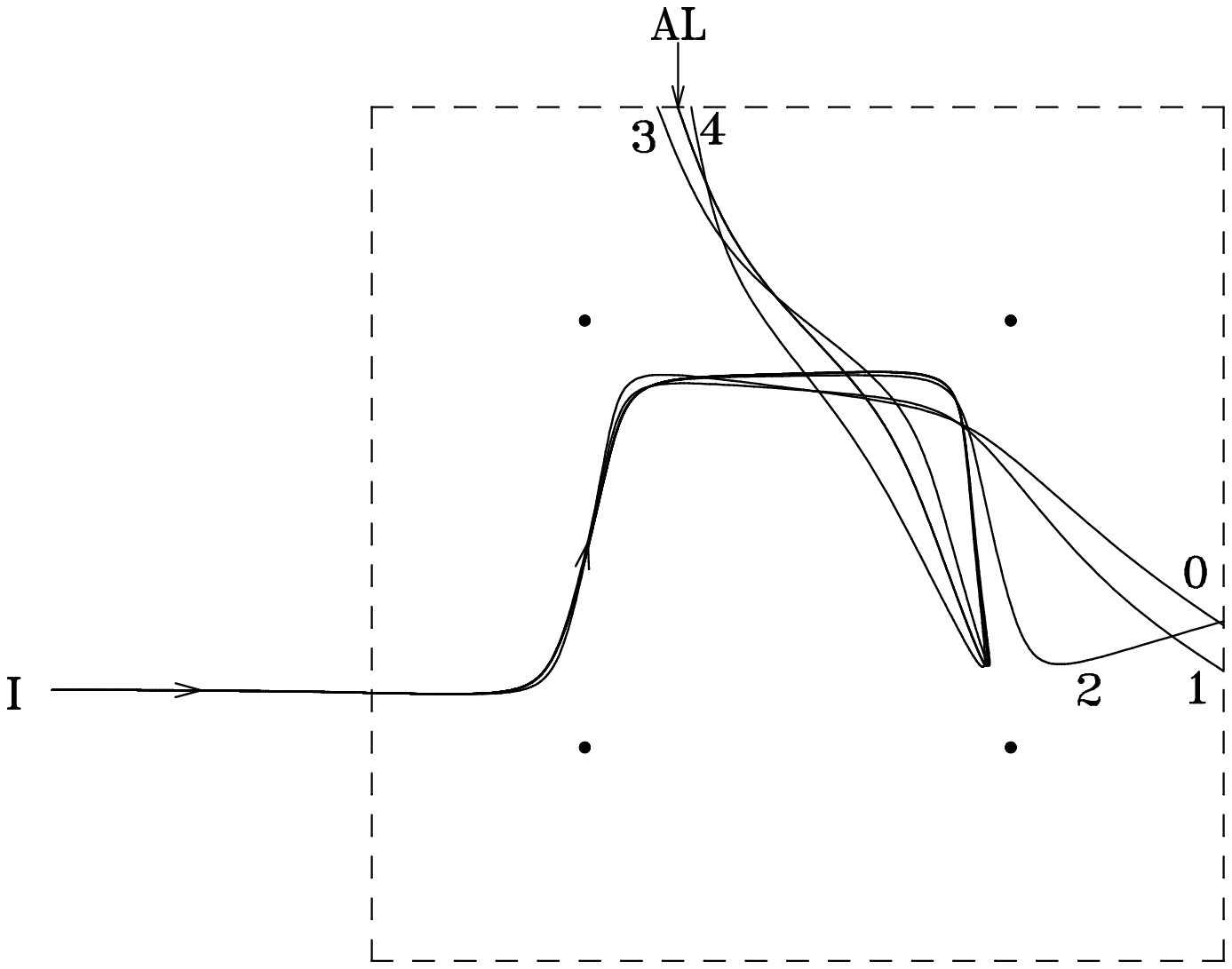}
\end{center}

\vspace{2cm}
Figure 2. Solutions of~(\ref{eq:AL}) corresponding to 
$\mathtt{maxiter} = 0,1,2,3,4$ with $\mathtt{accuracy} = 0$. The solution of
the regular order reduction of~(\ref{eq:AL}) and the
one obtained integrating it backwards are indistinguishable
and appear labelled as AL.

\newpage
\vspace*{2cm}
\begin{center}
\includegraphics[width=\textwidth]{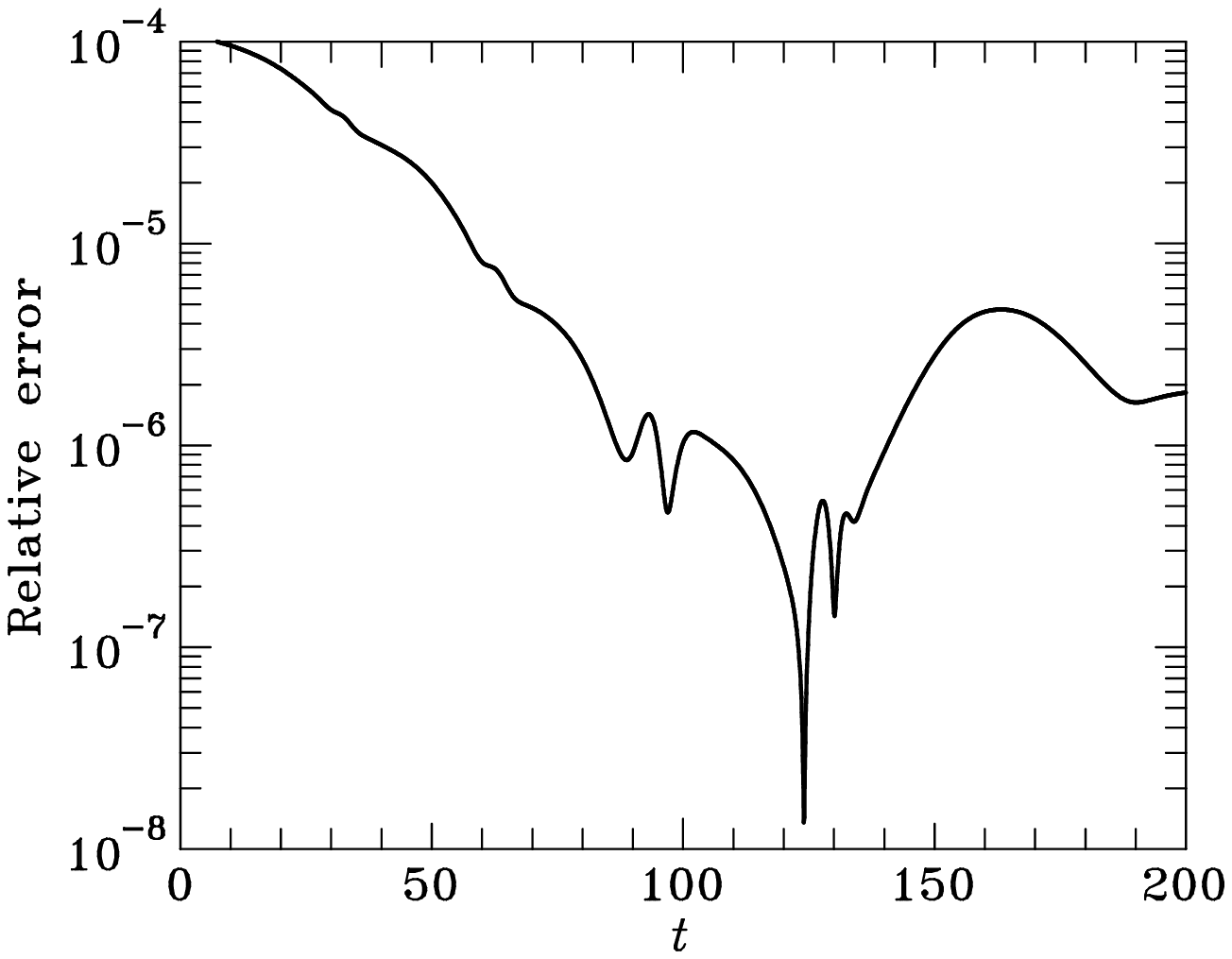}
\end{center}

\vspace{2cm}
Figure 3. Evolution of the relative distance between the solutions
obtained integrating forward and backward system~(\ref{eq:AL})
by means of \texttt{TestRed2.c}.

\newpage
\vspace*{2cm}
\begin{center}
\includegraphics[width=\textwidth]{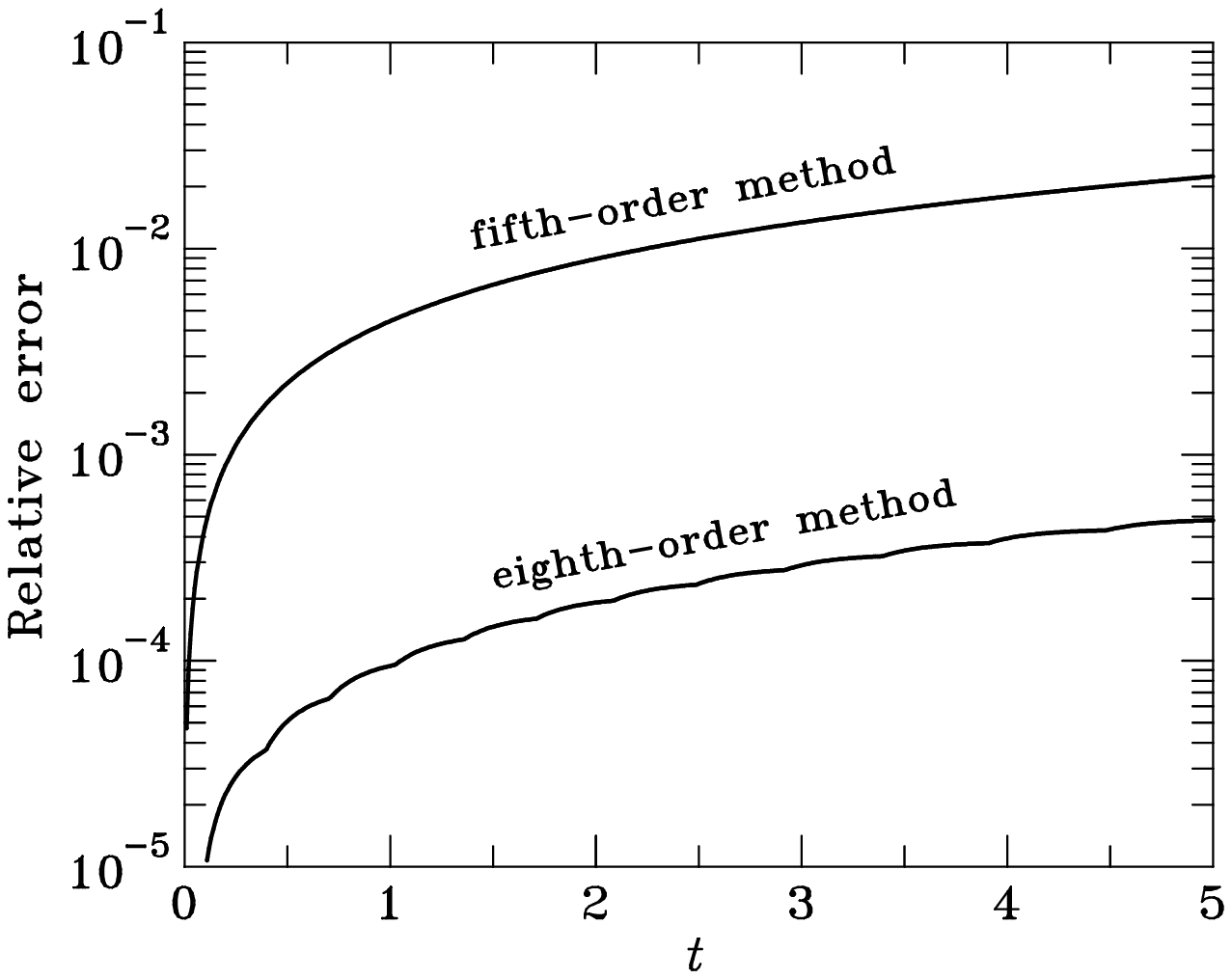}
\end{center}

\vspace{2cm}
Figure 4. Evolution of the accumulated error when computing
by means \texttt{TestRed3.c}
the regular order reduction~(\ref{eq:nodelay}) of~(\ref{eq:delay}).


\newpage
\begin{thebibliography}{99}

\bibitem{Rohrlich}
F.\ Rohrlich, Classical Charged Particles, 
(Addison-Wesley, Reading, MA, 1965).

\bibitem{Huschilt}
J.\ Huschilt and W.E.\ Baylis, Phys.\ Rev.\ D 13 (1976) 3256.

\bibitem{Baylis}
W.E.\ Baylis and J.\ Huschilt, Phys.\ Rev.\ D 13 (1976) 3262.

\bibitem{JMA}
J.M.\ Aguirregabiria, J.\ Phys.\ A 30 (1997) 2391.

\bibitem{Bhabha}
H.J.\ Bhabha, Phys.\ Rev.\ 70 (1946) 759.

\bibitem{Kerner}
E.\ Kerner, J.\ Math.\ Phys.\ 6 (1965) 1218.

\bibitem{Sanz}
J.L.\ Sanz, J.\ Math.\ Phys.\ 20 (1979) 2334.

\bibitem{Bel2}
Ll.\ Bel and X.\ Fustero, Ann.~Inst.~H.~Poincar\'e 25 (1976) 
411; and references therein.

\bibitem{Bel3}
Ll.\ Bel, in: Relativistic Action 
at a Distance: Classical and Quantum Aspects, ed.\ J.\ Llosa 
(Springer, Berlin, 1982), p.\ 21.

\bibitem{Bel4}
Ll.\ Bel, J.-L.\ Boulanger and N.\ Deruelle, Phys.\ Rev.\ A 37 
(1988) 1563.

\bibitem{Bel1}
Ll.\ Bel and H.\ Sirousse-Zia, Phys.\ Rev.\ D 32 (1985) 3128.

\bibitem{Parker}
L.\ Parker and J.Z.\ Simon, Phys.\ Rev.\ D 47 (1993) 1339.

\bibitem{sanedrin}
J.M.\ Aguirregabiria, A.\ Hern\'andez and M.\ Rivas,
J.\ Phys.\ A 30 (1997) L651--L654.

\bibitem{ODE}
J.M.\ Aguirregabiria, DS Workbench beta version. This unpublished 
program is a greatly improved version for Windows 3.1, 95 and NT 4.0
of: ODE Workbench, Physics Academic Software, (American Institute
of Physics, New York, 1994).

\bibitem{Bellman}
R.E.\ Bellman and K.L. Cooke, Differential-Difference Equations
(Academic Press, New York, 1963).

\bibitem{Hairer}
E.\ Hairer, S.P.\ Norsett and G.\ Wanner, Solving Ordinary
Differential Equations I: Nonstiff Problems, 2nd.\ ed.\ (Springer, 
Berlin, 1993).

\bibitem{lambert} R.M.\ Corless, G.H.\ Gonnet, D.E.G.\ Hare, D.J.\ Jeffrey
and D.E.\ Knuth, On the Lambert W Function, Technical Report CS-93-03,
Dept.\ Comp.\ Sci., University of Waterloo (1993).\\
\texttt{ftp://cs-archive.uwaterloo.ca/cs-archive/CS-93-03/W.ps.Z}.

\end{thebibliography}
\end{document}